\documentclass[a4paper,12pt]{report}
\usepackage[T2A]{fontenc}
\usepackage[cp1251]{inputenc}
\usepackage{amsfonts,amsmath,amssymb}
\usepackage[ukrainian,russian,english]{babel}
\usepackage{amsthm}
\usepackage{amscd}
\usepackage{amsmath}
\usepackage{amssymb}
\usepackage{graphicx}
\usepackage[T2A]{fontenc}
\usepackage[cp1251]{inputenc}
\def\beq{\begin{eqnarray}}
\def\eeq{\end{eqnarray}}

\theoremstyle{definition}

\theoremstyle{remark}

\begin{document}

{ Periodic solutions of Schrodinger equation in Hilbert space. \\
}

{\bf Boichuk A.A.$^{*}$, Pokutnyi A.A.$^{*}$}
%{\bf А.А.Pokutnyi$^{*}$, A.A.Boichuk$^{*}$, D.A.Nomirovsky$^{**}$ \\
%}

{\it * Institute of mathematics of NAS of Ukraine

%** Kiev National Taras Schevchenko University, faculty of
%cybernetics

e-mail: lenasas@gmail.com

e-mail: boichuk@imath.kiev.ua \\ }

Necessary and sufficient conditions for existence of boundary
value problem of Schrodinger equation are obtained in linear and
nonlinear cases. Periodic analytical solutions are
represented using generalized Green's operator. \\

{\bf Auxilary result(Linear case).}

{\bf Statement of the problem.} Consider the next boundary value
problem for Shrodinger equation  \beq \label{1}
\frac{d\varphi(t)}{dt} = -i H_{0}\varphi(t) + f(t), t \in [0;w]
\eeq \beq \label{2} \varphi(0) - \varphi(w) = \alpha \in D \eeq in
a Hilbert space $H_{T}$, where, for each
 $t \in [0;w]$, the unbounded operator $H_{0}$ has the form \cite{Rid2}
$$
H_{0} = i \left(\begin{array}{rcl} 0 &
T \\
- T & 0
\end{array}\right) = \left(\begin{array}{rcl} T  &
0 \\
0 & T
\end{array}\right)\left(\begin{array}{rcl} 0 &
 I \\
 -I & 0
\end{array}\right) = i \left(\begin{array}{rcl} 0 &
I \\
 -I & 0
\end{array}\right)\left(\begin{array}{rcl} T  &
0 \\
0 & T
\end{array}\right),
$$ for simplicity.
In more general case operator $H_{0}$ has the next form
$$
H_{0} = i J \left(\begin{array}{rcl} T  &
0 \\
0 & T
\end{array}\right) = i \left(\begin{array}{rcl} T  &
0 \\
0 & T
\end{array}\right) J, ~~~J = J^{*} = J^{-1},
$$
where $T$ is strongly positive self-adjoint operator in the
Hilbert space $H$. Since operator $T^{2}$ is closed, then domain
$D(T)$ of operator $T$ is Hilbert space with scalar product
$(Tu,Tu)$. The space $H_{T} = H \oplus H$ and operator $H_{0}$ is
self-adjoint on domain $D = D(T) \oplus D(T)$  with product
$$
(<u,v>, <u,v>)_{H^{T}} = (Tu,Tu)_{H} + (Tv,Tv)_{H}
$$
and infinitesemal generator of strongly continuous evolution
semigroup
$$
U(t): = U(t,0) = \left(\begin{array}{rcl} cos~tT &
sin~tT \\
-sin~tT & cos~tT
\end{array}\right),~~U^{n}(t) = \left(\begin{array}{rcl} cos~ntT^{\frac{1}{2}} &
sin~ntT^{\frac{1}{2}} \\
-sin~ntT^{\frac{1}{2}} & cos~ntT^{\frac{1}{2}}
\end{array}\right) = U(nt),
$$
$||U^{n}(t)|| = 1, n \in \mathbb{N}$ (nonexpanding group); $\varphi(t) = (\varphi_{1}(t), \varphi_{2}(t))^{T}$,
 $\alpha = (\alpha_{1}, \alpha_{2})^{T}$, $f(t) = (f_{1}(t), f_{2}(t))^{T} $. Solutions of equation
(\ref{1}) can be represented in the next form
$$
\varphi(t) = U(t)c + \int_{0}^{t}U(t)U^{-1}(\tau)f(\tau)d\tau,
$$
for any element $c \in H_{T}$. Substitute in condition (\ref{2})
we obtain that solvability of boundary value problem (\ref{1}),
(\ref{2}) is equivalent solvability the next operator equation
\beq \label{3} (I - U(w))c = g, \eeq where $g = \alpha + U(w)
\int_{0}^{w}U^{-1}(\tau)f(\tau)d\tau.$ Consider the case when the
set of values of $I - U(w)$ is closed $R(I - U(w)) = \overline{R(I
- U(w))}$.  Since $||U^{n}(w)|| = ||U(wn)|| = 1 $ for all $n \in
\mathbb{N}$ then \cite{BilBP} the operator system (\ref{3}) is
solvable if and only if \beq \label{4} U_{0}(w) g = 0, \eeq where
$$U_{0}(w) = \lim_{n \rightarrow \infty} \frac{\sum_{k =
0}^{n}U^{k}(w)}{n} =\lim_{n \rightarrow \infty} \frac{\sum_{k =
0}^{n}U(kw)}{n} -$$ orthoprojector, which projects the space
$H_{T}$ onto subspace $ 1 \in \sigma (U(w))$. Under this condition
solutions of (\ref{3}) have the form $$ c = U_{0}(w)\overline{c} +
(\sum_{k = 0}^{\infty}(\mu - 1)^{k} \{ \sum_{l = 0}^{\infty}
\mu^{-l-1}(U(w) - U_{0}(w))^{l} \}^{k + 1} - U_{0}(w))g,
$$
for $\mu > 1, |1 - \mu| < \frac{1}{||R_{\mu}(U(w))||}$ and any
$\overline{c} \in H_{T}$. Then we can formulate first result as
lemma.

{\bf Lemma 1. } {\it Let the operator $I - U(w)$ has a closed
image $R(I - U(w)) = \overline{R(I - U(w))}$.

1. There exist solutions of boundary value problem (\ref{1}),
(\ref{2}) if and only if \beq \label{5} U_{0}(w) ( \alpha +
\int_{0}^{w}U^{-1}(\tau)f(\tau)d\tau)  = 0. \eeq

2. Under condition (\ref{5}), solutions of (\ref{1}), (\ref{2}) have
the form \beq \label{6} \varphi(t, \overline{c}) =
U(t)U_{0}(w)\overline{c} + (G[f,\alpha])(t), \eeq where
$$
(G[f,\alpha])(t) = U(t)\sum_{k = 0}^{\infty}(\mu - 1)^{k} \{ \sum_{l
= 0}^{\infty} \mu^{-l-1}(U(w) - U_{0}(w))^{l} \}^{k + 1} (\alpha +
\int_{0}^{w} U(w)U^{-1}(\tau)f(\tau)d\tau ) -
$$
$$
- U(t)U_{0}(w)(\alpha + \int_{0}^{w}U(w)U^{-1}(\tau)f(\tau)d\tau) +
\int_{0}^{t}U(t)U^{-1}(\tau)f(\tau)d\tau,
$$
is the generalized Green operator of the boundary value problem
(\ref{1}), (\ref{2}) for $\mu > 1, |1 - \mu| <
1/||R_{\mu}(U(w))||$.}

Now we show that condition $R(I - U(w)) = \overline{R(I - U(w))}$
of lemma 1 can be omitted and in the different senses boundary
value problem (\ref{1}), (\ref{2}) is always resolvable.

 1) Classical generalized solutions.

Consider case when the set of values of $I - U(w)$ is closed ($R(I -
U(w)) = \overline{R(I - U(w))}$). Then \cite{BoiSam} $g \in R(I -
U(w))$ if and only if $\mathcal{P}_{N(I - U(w))^{*})}g = 0$ and the
set of solutions of (\ref{3}) has the form \cite{BoiSam} $c = G[g] +
U_{0}(w)\overline{c}, \forall \overline{c} \in H_{T}, $ where
\cite{BilBP} and \cite{BoiSam}
$$
G[g] = (I - U(w))^{+}g = ((I - (U(w) - U_{0}(w))^{-1} - U_{0}(w))g
$$
is generalized Green operator (or in the form of convergent series).

2) Strong generalized solutions. Consider the case when $R(I - U(w))
\neq \overline{R(I - U(w))}$ and $g \in \overline{R(I - U(w))}$. We
show that operator $I - U(w)$ may be extended to $\overline{I -
U(w)}$ in such way that $R(\overline{I - U(w)})$ is closed.

Since the operator $I - U(w)$ is bounded the next representation of
$H_{T}$ in the direct sum is true
$$
H_{T} = N(I - U(w)) \oplus X, H_{T} = \overline{R(I - U(w))} \oplus
Y,
$$
with $X = N(I - U(w))^{\bot} = \overline{R(I - U(w))}$ and $Y =
\overline{R(I - U(w))}^{\bot} = N(I - U(w))$. Let $E = H_{T}/N(I -
U(w))$ is quotient space of $H_{T}$ , $\mathcal{P}_{\overline{R(I -
U(w))}}$ and $\mathcal{P}_{N(I - U(w))}$   are orthoprojectors,
which project onto $\overline{R(I - U(w))}$ and $N(I - U(w))$
respectively. Then operator
$$
\mathcal{I - U}(w) = \mathcal{P}_{\overline{R(I - U(w))}}(I -
U(w))j^{-1}p : X \rightarrow R(I - U(w)) \subset \overline{R(I -
U(w))},
$$
is linear, continuous and injective. Here
$$
p : X \rightarrow E = H_{T}/N(I - U(w)), ~~ j : H_{T} \rightarrow E
$$ are continuous bijection and projection respectively. The triple $(H_{T}, E, j)$ is a locally trivial bundle
with typical fiber $H_{1} = \mathcal{P}_{N(I - U(w))}H$
\cite{Atiyah}.  In this case \cite[p.26,29]{Lash} we can define
strong generalized solution of equation \beq \label{16} (\mathcal{I
- U}(w)) x = g, x \in X . \eeq Fill up the space $X$ in the norm
$||x||_{\overline{X}} = ||(\mathcal{I - U}(w))x||_{F},$ where $F =
\overline{R(I - U(w))}$ \cite{Lash}. Then extended operator
$$\overline{\mathcal{I - U}(w)} : \overline{X} \rightarrow \overline{R(I - U(w))}, X \subset \overline{X}$$
is homeomorphism of $\overline{X}$ and $\overline{R(I - U(w))}$. By
virtue of construction of strong generalized solution \cite{Lash}
equation
$$
(\overline{\mathcal{I - U}(w)}) \overline{\xi} = g,
$$
has a unique solution $(\overline{\mathcal{I - U}(w)})^{-1}g$ which
is called generalized solution of equation (\ref{16}).

{\bf Remark 1.} It should be noted that there are exists next
extensions of spaces and corresponding operators
$$
\overline{p} : \overline{X} \rightarrow \overline{E},~~ \overline{j}
: \overline{H}_{T} \rightarrow \overline{E},~~
\overline{\mathcal{P}_{X}} = \mathcal{P}_{\overline{X}} :
\overline{H}_{T} \rightarrow \overline{X},~~ \overline{G} :
\overline{R(I - U(w))} \rightarrow \overline{X},
$$
where
$$\overline{H}_{T} = N(I - U(w)) \oplus \overline{X};~~\overline{p}(x) = p(x), x \in X; ~~\overline{j}(x) = j(x), x \in
H_{T}, $$ $$\overline{\mathcal{P}}_{X}(x) = \mathcal{P}_{X}(x), x
\in H_{T} ~(\mathcal{P}_{X} = \mathcal{P}^{2}_{X} =
\mathcal{P}^{*}_{X});~~\overline{G}[g] = G[g], g \in R(I - U(w)).$$

Then the operator $\overline{I - U(w)} = (\overline{\mathcal{I -
U}(w)})\mathcal{P}_{\overline{X}} : \overline{H}_{T} \rightarrow
\overline{H}_{T}$ is extension of $I - U(w)$, $\overline{(I -
U(w))}c = (I - U(w))c$ for all $c\in H_{T}$ .

3) Strong pseudosolutions.

Consider element $g \notin \overline{R(I - U(w))}$. This condition
is equivalent $\mathcal{P}_{N(I - U(w))^{*})}g \neq 0$. In this case
there are exists elements from $\overline{H_{T}}$ which minimise
norm $||\overline{(I - U(w))}\xi - g ||_{\overline{H_{T}}}$ :
$$
\xi = \overline{(\mathcal{I - U}(w))}^{-1}g + \mathcal{P}_{N(I -
U(w))}\overline{c}, \forall \overline{c} \in H_{T}.
$$
These elements we call {\it strong pseudosolutions} by analogy of
\cite{BoiSam}.

Now we formulate the full theorem of solvability.

{\bf Theorem 1.} {\it Boundary value problem (\ref{1}), (\ref{2}) is
always resolvable.

1) a) There are exists classical or strong generalized solutions of
(\ref{1}), (\ref{2}) if and only if \beq \label{17} U_{0}(w) (
\alpha + \int_{0}^{w}U^{-1}(\tau)f(\tau)d\tau)  = 0. \eeq If
$(\alpha + \int_{0}^{w}U^{-1}(\tau)f(\tau)d\tau) \in R(I - U(w))$
then solutions of  (\ref{1}), (\ref{2}) will be classical.

b) Under assumption (\ref{17}) solutions of (\ref{1}), (\ref{2})
have the form $$ \varphi(t, \overline{c}) = U(t)U_{0}(w)\overline{c}
+ (\overline{G[f,\alpha]})(t),$$ where $(\overline{G[f,\alpha]})(t)$
- is extension of operator $(G[f,\alpha])(t)$; \\

3) a) There are exists strong pseudosolutions if and only if \beq
\label{18} U_{0}(w) ( \alpha + \int_{0}^{w}U^{-1}(\tau)f(\tau)d\tau)
\neq 0. \eeq

b) Under assumption (\ref{18})  strong pseudosolutions of (\ref{1}),
(\ref{2}) have the form $$ \varphi(t, \overline{c}) =
U(t)U_{0}(w)\overline{c} + (\overline{G[f,\alpha]})(t), $$ where
$$
(\overline{G[f,\alpha]})(t) = U(t) \overline{G}[g] +
\int_{0}^{t}U(t)U^{-1}(\tau)f(\tau)d\tau = U(t)\overline{(\mathcal{I
- U}(w))}^{-1}g + \int_{0}^{t}U(t)U^{-1}(\tau)f(\tau)d\tau.
$$
 }

{\bf Main result (Nonlinear case). Generalization of
Lyapunov-Schmidt method}

In the Hilbert space $H_{T}$ defined below we consider the
boundary value problem \beq \label{7} \frac{d\varphi(t)}{dt} =
-iH_{0}\varphi(t) + \varepsilon Z(\varphi(t),t,\varepsilon) +
f(t), \eeq \beq \label{8} \varphi(0, \varepsilon) - \varphi(w,
\varepsilon) = \alpha. \eeq  We seek a bounded solution
$\varphi(t,\varepsilon)$ of boundary value problem (\ref{7}),
(\ref{8}) that becomes one of the solutions of the generating
equation (\ref{1}), (\ref{2}) $\varphi_{0}(t, \overline{c})$ in
the form (\ref{6}) for $\varepsilon = 0$.

To find a necessary condition on the operator function $Z(\varphi,
t, \varepsilon),$ we impose the joint constraints
$$
Z(\cdot, \cdot, \cdot) \in C([0; w],H_{T})\times
C[0,\varepsilon_{0}]\times C[||\varphi - \varphi_{0}||\leq q],
$$
where $q$ is some positive constant.

The main idea of the next results is presented in \cite{BoiPok} for investigating of bounded solutions.

Let us show that this problem can be solved with the use of the
operator equation for generating amplitudes \beq \label{9}
F(\overline{c}) = U_{0}(w) \int_{0}^{w}
U^{-1}(\tau)Z(\varphi_{0}(\tau, \overline{c}), \tau, 0)d\tau = 0.
\eeq

{\bf Theorem 2. }(necessary condition) {\it Let the nonlinear
boundary value problem (\ref{7}), (\ref{8}) has a bounded solution
$\varphi(\cdot, \varepsilon)$ that becomes one of the solutions
$\varphi_{0}(t, \overline{c})$ of the generating equation
(\ref{1}), (\ref{2}) with constant $\overline{c} = c^{0},$
$\varphi(t, 0) = \varphi_{0}(t, c^{0})$ for $\varepsilon = 0$.
Then this constant should satisfy the equation for generating
amplitudes (\ref{9}).}

To find a sufficient condition for the existence of solutions of
boundary value problem (\ref{7}), (\ref{8}) we additionally assume
that the operator function $Z(\varphi, t, \varepsilon)$ is
strongly differentiable in a neighborhood of the generating
solution $(Z(\cdot, t, \varepsilon) \in C^{1}[||\varphi -
\varphi_{0}|| \leq q])$.

This problem can be solved with the use of the operator
$$
B_{0} = \frac{dF(\overline{c})}{d\overline{c}}|_{\overline{c} =
c_{0}} =
\int_{-\infty}^{+\infty}H(t)A_{1}(t)T(t,0)P_{+}(0)\mathcal{P}_{N(D)}dt
: H \rightarrow H,
$$
where $A_{1}(t) = Z^{1}(v, t, \varepsilon)|_{v = \varphi_{0},
\varepsilon = 0}$ (the Fr\'{e}chet derivative).

{\bf Theorem 3.} (sufficient condition) {\it
 Let the operator $B_{0}$ satisfy the following conditions:

1) The operator $B_{0}$ is Moore-Penrouse pseudoinvertible;

2) $\mathcal P_{N(B_{0}^{*})} U(w) = 0.$

Then for arbitrary element $c = c^{0} \in H_{T}$, satisfying the
equation for generating amplitudes (\ref{9}), there exists at
least one solution of (\ref{7}), (\ref{8}).

This solution can be found with the use of the iterative process:
$$
\overline{v}_{k+1}(t, \varepsilon) = \varepsilon
G[Z(\varphi_{0}(\tau, c^{0}) + v_{k}, \tau , \varepsilon),
\alpha](t),
$$
$$
c_{k} = - B_{0}^{+}U_{0}(w)\int_{0}^{w}
U^{-1}(\tau)\{A_{1}(\tau)\overline{v}_{k}(\tau,\varepsilon) +
\mathcal R (v_{k}(\tau,\varepsilon), \tau, \varepsilon) \}d\tau,
$$
$$
 v_{k+1}(t, \varepsilon) = U(t)U_{0}(w)c_{k} +
\overline{v}_{k+1}(t, \varepsilon),
$$
$$
\varphi_{k}(t, \varepsilon) = \varphi_{0}(t, c^{0}) + v_{k}(t,
\varepsilon), k = 0, 1, 2, ..., ~~v_{0}(t, \varepsilon) =
0,\varphi(t, \varepsilon) = \lim_{k \rightarrow \infty}
\varphi_{k}(t, \varepsilon).
$$
} {\bf Remark 2.} Proof of theorems 2 and 3 follows directly from
works \cite{BoiPok}, \cite{Pok}. \\

{\bf Relationship between necessary and sufficient conditions.}

First, we formulate the following assertion.

{\bf Corollary.} {\it Let a functional $F(\overline{c})$ have the
Fr\'{e}chet derivative $F^{(1)}(\overline{c})$ for each element
$c^{0}$ of the Hilbert space $H$ satisfying the equation for
generating constants (\ref{9}). If $F^{1}(\overline{c})$ has a
bounded inverse, then boundary value problem (\ref{7}), (\ref{8})
has a unique solution for each
$c^{0}$.}

{\bf Remark 3.} If assumptions of the corollary are satisfied,
then it follows from its proof that the operators $B_{0}$ and
$F^{(1)}(c^{0})$ are equal. Since the operator
$F^{(1)}(\overline{c})$ is invertible, it follows that assumptions
1 and 2 of Theorem 3 are necessarily satisfied for the operator
$B_{0}$. In this case, boundary value problem (\ref{7}), (\ref{8})
has a unique bounded solution for each $c^{0} \in H_{T}$
satisfying (\ref{9}). Therefore, the invertibility condition for
the operator $F^{1}(\overline{c})$ relates the necessary and
sufficient conditions. In the finite-dimensional case, the
condition of invertibility of the operator $F^{(1)}(\overline{c})$
is equivalent to the condition of simplicity of the root $c^{0}$
of the equation for generating amplitudes \cite{BoiSam}.

 In such way we generalize the well-known method of Lyapunov-Schmidt. It should be emphasized that theorem 2 and 3 give us condition of chaotic behavior of (\ref{7}),
 (\ref{8}) \cite{Chuesh}.

 {\bf Example.} Now we illustrate obtained assertion.  Consider
 the next differential equation in separable Hilbert space $H$

 \beq \label{11} \ddot{y}(t) + T y(t) = \varepsilon (1 -
||y(t)||^{2})\dot{y}(t), \eeq \beq \label{12} y(0) = y(w),
~~\dot{y}(0) = \dot{y}(w), \eeq where $T$ is unbounded operator
with compact $T^{-1}$. Then there is exists orthonormal basis
$e_{i} \in H$ such that $y(t) = \sum_{i =
1}^{\infty}c_{i}(t)e_{i}$ and $Ty(t) = \sum_{i =
1}^{\infty}\lambda_{i}c_{i}(t)e_{i},$ $\lambda_{i} \rightarrow
\infty$.  Operator system (\ref{7}), (\ref{8}) for boundary value
problem (\ref{11}), (\ref{12}) in this case will be equivalent the
next countable system of ordinary differential equations
$(c_{k}(t) = x_{k}(t))$
$$ \dot{x}_{k}(t) =
\sqrt{\lambda_{k}}y_{k}(t),~~k = 1, 2, ...,$$ \beq \label{14}
\dot{y}_{k}(t) = -\sqrt{\lambda_{k}}x_{k}(t) + \varepsilon
\sqrt{\lambda_{k}}(1 - \sum_{j = 1}^{\infty}x_{j}^{2}(t))y_{k}(t),
\eeq \beq \label{15} x_{k}(0) = x_{k}(w), y_{k}(0)=y_{k}(w). \eeq
We find solutions of these equations such that for $\varepsilon =
0$ turns in one of the solutions of generating equation. Consider
critical case $\lambda_{i} = 4\pi^{2}i^{2}/w^{2}, i \in
\mathbb{N}$. Let $w = 2\pi$. In that case the set of all periodic
solutions of (\ref{14}), (\ref{15}) have the form
$$
x_{k}(t) = cos(kt)c_{1}^{k} + sin(kt)c_{2}^{k},
$$
$$
y_{k}(t) = -sin(kt)c_{1}^{k} + cos(kt)c_{2}^{k},
$$
for all pairs of constant $c_{1}^{k}, c_{2}^{k} \in \mathbb{R}, k
\in \mathbb{N}$. Equation for generating amplitudes (\ref{9}) in
this case will be equivalent the next countable systems of
algebraic nonlinear equations
$$
(c_{1}^{k})^{3} + 2 \sum_{j = 1, j \neq
k}(c_{1}^{k}(c_{1}^{j})^{2} + c_{1}^{k}(c_{2}^{j})^{2}) +
c_{1}^{k}(c_{2}^{k})^{2} - 4c_{1}^{k} = 0,
$$
$$
(c_{2}^{k})^{3} + 2 \sum_{j = 1, j \neq
k}(c_{2}^{k}(c_{1}^{j})^{2} + c_{2}^{k}(c_{2}^{j})^{2}) +
(c_{1}^{k})^{2}c_{2}^{k} - 4c_{2}^{k} = 0, k \in \mathbb{N}.
$$
Then we can obtain the next result

{\bf Theorem 4}(necessary condition of van der Pol's equation).{\it
Let the boundary value (\ref{14}), (\ref{15}) have a bounded
solution $\varphi(\cdot, \varepsilon)$ that becomes one of the
solutions of the generating equations with pairs of constant
$(c_{1}^{k}, c_{2}^{k}), k \in \mathbb{N}$. Then only finite number
of these pairs are not equal zero. Moreover,  if $(c_{1}^{k_{i}},
c_{2}^{k_{i}}) \neq (0, 0), i = \overline{1,N}$ then these constants
lie on N-dimensional torus in infinite dimensional space of
constants
$$
(c_{1}^{k_{i}})^{2} + (c_{2}^{k_{i}})^{2} =
(\frac{2}{\sqrt{2N-1}})^{2}, i = \overline{1,N}.
$$  }

\end{document}